\documentclass[12pt,epsf]{article}
\usepackage{graphicx}
\usepackage{amsmath,amssymb}
\begin{document}

\def\a{\alpha}
\def\b{\beta}
\def\c{\varepsilon}
\def\d{\delta}
\def\e{\epsilon}
\def\f{\phi}
\def\g{\gamma}
\def\h{\theta}
\def\k{\kappa}
\def\l{\lambda}
\def\m{\mu}
\def\n{\nu}
\def\p{\psi}
\def\q{\partial}
\def\r{\rho}
\def\s{\sigma}
\def\t{\tau}
\def\u{\upsilon}
\def\v{\varphi}
\def\w{\omega}
\def\x{\xi}
\def\y{\eta}
\def\z{\zeta}
\def\D{\Delta}
\def\G{\Gamma}
\def\H{\Theta}
\def\L{\Lambda}
\def\F{\Phi}
\def\P{\Psi}
\def\S{\Sigma}
\def\tto{\,-\,}

\def\o{\over}
\def\beq{\begin{eqnarray}}
\def\eeq{\end{eqnarray}}
\newcommand{\gsim}{ \mathop{}_{\textstyle \sim}^{\textstyle >} }
\newcommand{\lsim}{ \mathop{}_{\textstyle \sim}^{\textstyle <} }
\newcommand{\vev}[1]{ \left\langle {#1} \right\rangle }
\newcommand{\bra}[1]{ \langle {#1} | }
\newcommand{\ket}[1]{ | {#1} \rangle }
\newcommand{\EV}{ {\rm eV} }
\newcommand{\KEV}{ {\rm keV} }
\newcommand{\MEV}{ {\rm MeV} }
\newcommand{\GEV}{ {\rm GeV} }
\newcommand{\TEV}{ {\rm TeV} }
\def\diag{\mathop{\rm diag}\nolimits}
\def\Spin{\mathop{\rm Spin}}
\def\SO{\mathop{\rm SO}}
\def\O{\mathop{\rm O}}
\def\SU{\mathop{\rm SU}}
\def\U{\mathop{\rm U}}
\def\Sp{\mathop{\rm Sp}}
\def\SL{\mathop{\rm SL}}
\def\tr{\mathop{\rm tr}}

\def\IJMP{Int.~J.~Mod.~Phys. }
\def\MPL{Mod.~Phys.~Lett. }
\def\NP{Nucl.~Phys. }
\def\PL{Phys.~Lett. }
\def\PR{Phys.~Rev. }
\def\PRL{Phys.~Rev.~Lett. }
\def\PTP{Prog.~Theor.~Phys. }
\def\ZP{Z.~Phys. }


\baselineskip 0.7cm

\begin{titlepage}

\begin{flushright}
IPMU 15-0066\\
\end{flushright}

\vskip 1.35cm
\begin{center}
{\large \bf
Muon $g-2$ in Focus Point SUSY
}
\vskip 1.2cm
  Keisuke Harigaya$^{a}$,  Tsutomu T. Yanagida$^{b}$ and Norimi Yokozaki$^{c}$
\vskip 0.4cm

{\it
$^a$ ICRR, the University of Tokyo, Kashiwa, 277-8582, Japan \\
$^b$ Kavli IPMU (WPI), UTIAS, the University of Tokyo, Kashiwa, 277-8583, Japan \\
$^c$ INFN, Sezione di Roma, Piazzale A. Moro 2, I-00185 Roma, Italy
}

\vskip 1.5cm

\abstract{
We point out that the anomaly of the muon $g-2$ can be easily explained in a focus point supersymmetry scenario,
which realizes the semi-natural supersymmetry.
Among known focus point supersymmetry scenarios,
we find that a model based on Higgs-gaugino mediation works with a mild fine-tuning 
$\Delta=40$\,-\,$80$.
We propose two new focus point supersymmetry scenarios where the anomaly of the muon $g-2$ is also explained.
These scenarios are variants of the widely known focus point supersymmetry based on gravity mediation
with universal scalar masses.
}

\vspace{20pt}
\end{center}

\end{titlepage}

\setcounter{page}{2}

\section{Introduction}
Low-energy supersymmetry (SUSY) has many attractive features, and is a leading candidate for physics beyond the standard model (SM).
In the minimal supersymmetric SM (MSSM),  three gauge coupling constants of SM gauge groups are unified at a high energy scale around $10^{16}$ GeV.
The electroweak symmetry breaking (EWSB) is induced via SUSY breaking,
which was expected to solve the fine-tuning problem of the Higgs potential,
namely, to explain the smallness of the EWSB breaking scale.

Another attractive and important feature of low-energy SUSY is, that it has a potential of providing a solution to the long-standing puzzle, the anomaly of the muon anomalous magnetic moment ($g-2$). The experimental value of the muon anomalous magnetic moment is deviated from the SM prediction $(a_\mu)_{\rm SM}$ above 3$\sigma$ level:
\begin{eqnarray}
(a_\mu)_{\rm EXP} - (a_\mu)_{\rm SM} = 
\left\{
\begin{array}{c}
  ( 26.1  \pm 8.0)\times 10^{-10} ~\cite{gm2_hagiwara} \\
  ( 28.7 \pm 8.0 )\times 10^{-10}  ~\cite{gm2_davier} \\
\end{array}
\right\}. \label{eq:gm2_dis}
\end{eqnarray}
Here, $(a_\mu)_{\rm EXP}$ is the experimental value of the muon $(g-2)/2$
accurately measured at the Brookhaven E821 experiment~\cite{gm2_exp}. 
In low-energy SUSY, smuons and chargino/neutralino of $\mathcal O$(100) GeV give $\mathcal{O}(10^{-9})$ corrections to the muon $g-2$ and explain this discrepancy~\cite{gm2_susy1,gm2_susy2}.

However, non-observation of SUSY signals at the Large Hadron Collider (LHC) (see e.g.~Refs.~\cite{lhc_susy})
and the relatively heavy Higgs boson of 125 GeV~\cite{lhc_higgs} push up the SUSY scale above TeV. 
Especially, the observed Higgs boson mass requires rather large radiative corrections from heavy stops~\cite{higgs_rad}:
it is suggested that the stop is as heavy as 3\,-\,5 TeV~\cite{higgs_3loop},
including higher order corrections beyond the 3-loop level.
As a result, both the SUSY solution to the fine-tuning problem and the SUSY explanation of the muon $g-2$ anomaly seem to be difficult to work.

There are several attempts to attack these two difficulties, but separately. As a solution to the fine-tuning problem, the focus point SUSY now becomes more attractive~\cite{fp_original} (see also~\cite{feng_new, Brummer:2013dya}).
In the focus point SUSY,  
a special relation among soft SUSY breaking parameters is assumed so that radiative corrections to the Higgs potential cancel each other.
As a result, the EWSB scale becomes insensitive to the soft SUSY breaking mass scale. 
There are several focus point SUSY scenarios, based on
gaugino mediation~\cite{fp_gaugino},
Higgs-gaugino mediation~\cite{fp_hyy},
gravity mediation with non-universal gaugino masses~\cite{fp_grav_nu_gaugino}
and gauge mediation~\cite{fp_gauge}.

On the other hand,
light smuons and light chargino/neutralino are required to explain the muon $g-2$ anomaly, while the Higgs boson mass around 125 GeV requires rather heavy stops. In Refs.~\cite{split_gm2}, it is shown that the Higgs boson mass and the muon $g-2$ anomaly are explained simultaneously by mass-splitting among generations. 
Also, other possibilities are provided based on gauge mediation~\cite{gm2_gauge}, gravity mediation~\cite{gm2_gravity} and gaugino mediation~\cite{gm2_gaugino}: in these frameworks, colored and non-colored SUSY particles are split in their masses so that the SUSY contribution to the muon $g-2$ is enhanced.

In this paper, we show that
the anomaly of the muon $g-2$ can be easily explained in a focus point SUSY scenario.
In the next section, we review four known types of focus point scenarios and
discuss whether the scenarios can explain the muon $g-2$ anomaly.
We find that a model based on Higgs-gaugino mediation,
which is recently proposed by the current authors~\cite{fp_hyy},
works.
It is found that the discrepancy of the muon $g-2$ from the SM prediction is reduced to 1$\sigma$ level with a mild fine-tuning $\Delta=40$\,-\,$80$.
(See Eq. (\ref{eq:ft_hgm}) for the definition of $\Delta$.)
We propose two new focus point scenarios which can explain the muon $g-2$ anomaly in section~\ref{sec:fpnus}.
They are variants of the well-known focus point SUSY scenario proposed by Feng, Matchev and Moroi~\cite{fp_original}.

\section{Focus point for the electroweak symmetry breaking}
\label{sec:focus EWSB}

In focus point SUSY scenarios, the EWSB scale
is relatively insensitive to the soft SUSY breaking mass scale. 
This is achieved by introducing some fixed ratios between soft mass parameters at a high energy scale. 
In this section, we review four known focus point scenarios
and discuss whether they can explain  the muon $g-2$ anomaly.
We show that only one of them
works.

The conditions for the EWSB are given by
\begin{eqnarray}
\frac{g_1^2 + g_2^2}{4} v^2 &\simeq& \left[ -\mu^2 
- \frac{(m_{H_u}^2  + \frac{1}{2 v_u}\frac{\partial \Delta V}{\partial v_u} ) \tan^2\beta}{\tan^2\beta-1} 
 + \, \frac{m_{H_d}^2 + \frac{1}{2 v_d}\frac{\partial \Delta V}{\partial v_d} }{\tan^2\beta-1} \right] \Bigr|_{M_{\rm IR}}, \nonumber \\
\frac{B \mu \,(\tan^2\beta+1)}{\tan\beta} &\simeq& \left[ m_{H_u}^2 +\frac{1}{2 v_u}\frac{\partial \Delta V}{\partial v_u} + m_{H_d}^2  + \frac{1}{2 v_d}\frac{\partial \Delta V}{\partial v_d} + 2\mu^2 \right] \Bigr|_{M_{\rm IR}}, \label{eq:ewsb}
\end{eqnarray}
where $g_1$ and $g_2$ are gauge coupling constants of $U(1)_Y$ and $SU(2)_L$,
$v_u$ and $v_d$ are the vacuum expectation values of the up-type and down-type Higgs,
${\rm tan}\beta \equiv v_u/v_d$,
$v\equiv \sqrt {v_u^2 + v_d^2}$ is the EWSB scale,
$\mu$ is the Dirac mass term of the Higgs doublets,
$m_{H_u}$ and $m_{H_d}$ are soft masses for the up-type and down-type Higgses,
$B \mu$ is the SUSY breaking holomorphic Higgs quadratic mass term,
and $\Delta V$ is a one-loop contribution to the Higgs potential.
$m_{H_u}^2$, $m_{H_d}^2$ and $\Delta V$ are evaluated at the geometric mean value of stop masses, $M_{\rm IR}=\sqrt{m_{Q_3} m_{ \bar{U}_3 }}$.
For large $\tan\beta$, $m_{H_d}^2$ is relatively unimportant for the EWSB scale, since its effect is suppressed by $1/\tan^2\beta$. 

The low-energy value of $m_{H_u}^2$ and $m_{H_d}^2$ are written in terms of gaugino masses and scalar masses at the high energy scale:%
\footnote{
Here, we neglect the contribution from A-terms, for simplicity.
It does not change our conclusion qualitatively unless A-terms are so large that they dominate the quantum corrections to $m_{H_u}^2$.
}
\begin{eqnarray}
m_{H_u}^2 (3\,{\rm TeV}) &\simeq& 0.009 M_1^2 + 0.217 M_2^2 - 1.168 M_3^2 \nonumber \\
&+& 0.005 M_1 M_2  -0.109 M_2 M_3  - 0.016 M_1 M_3 \nonumber \\
&+&  0.667 m_{H_u}^2 + 0.026 m_{H_d}^2 +  0.073 m_L^2 -  0.074 m_{\bar E}^2 \nonumber \\
&-&  0.385 m_Q^2 -  0.163 m_{\bar U}^2 - 0.070 m_{\bar D}^2, \nonumber \\
%
%
m_{H_d}^2 (3\,{\rm TeV}) &\simeq& 
0.030 M_1^2 + 0.367 M_2^2 - 0.120M_3^2 \nonumber \\
&-& 0.002 M_1 M_2  -0.030 M_2 M_3  - 0.001M_1 M_3 \nonumber \\
&+&  0.019 m_{H_u}^2 + 0.933 m_{H_d}^2  -0.088 m_L^2 +  0.063 m_{\bar E}^2 \nonumber \\
&+&  0.044 m_Q^2 -  0.145 m_{\bar U}^2 +0.043 m_{\bar D}^2, 
\label{eq:mhu2_msusy}
\end{eqnarray}
for $M_{\rm IR}=3$ TeV, $\tan\beta=20$, $m_{t}=173.34$ GeV and  $\alpha_s(m_Z)=0.1185$. 
The soft SUSY breaking parameters in the right hand side of Eq.~(\ref{eq:mhu2_msusy}) are defined at $M_{\rm in}=10^{16}$~GeV. 
Here, $M_1$, $M_2$ and $M_3$ are the bino, wino and gluino masses,
respectively,
and
$m_Q$, $m_{\bar{U}}$, $m_{\bar{D}}$, $m_L$ and $m_{\bar{E}}$ are
generation-universal
soft masses of
left-handed squarks, right-handed up squarks, right-handed down squarks, left-handed sleptons and right-handed sleptons, respectively.
The above expressions are obtained by numerically solving two-loop renormalization group equations~\cite{Martin:1993zk}. For this purpose, we use {\tt SoftSUSY 3.6.1} package~\cite{softsusy}.

In the focus point SUSY, $\tilde m_{H}^2 \equiv m_{H_u}^2-(m_{H_d}^2-m_{H_u}^2)/\tan^2\beta$ becomes insensitive to SUSY breaking parameters. This is achieved by introducing fixed ratio(s) among mass parameters at a high energy scale,
which we take as $M_{\rm in}= 10^{16}$~GeV.
Currently, the following four focus point scenarios are known.

\begin{description}
\item[FPUS] : universal scalar masses ($m_0$) and a fixed  $m_0/M_3$ 
\item[FPGM] : vanishing or small scalar masses and a fixed  $M_2/M_3$
\item[FPHSG] : high scale gauge mediation with a fixed messenger number, $(N_2, N_3)$
\item[FPHGM] : vanishing slepton and squark masses and a fixed $m_{H_u}/M_3$
\end{description}

{\bf FPUS} is based on gravity mediation, where
Universal Scalar masses are assumed. In this case, their contributions to $m_{H_u}^2$ almost cancel each other.
{\bf FPGM} is based on Gaugino Mediation, where all soft scalar masses vanish at the high energy scale $M_{\rm in}$.
{\bf FPHSG} is based on High Scale Gauge mediation, where scalar masses as well as gaugino masses are generated by messenger loops. 
Finally, {\bf FPHGM} is based on Higgs-Gaugino Mediation motivated by $E_7$ non-linear sigma model~\cite{Kugo:1983ai}, where squark and slepton masses vanish at the high energy scale.
More detailed descriptions are shown below.

\vspace{10pt}
Before discussing each focus point, we comment on non-universal gaugino masses.
As we will see in the next section, non-universal gaugino masses are crucial in order to explain the muon $g-2$ anomaly and the observed Higgs boson mass around 125 GeV, simultaneously.
Non-universal gaugino masses are naturally obtained if product group unification (PGU) is considered~\cite{ArkaniHamed:1996jq}.
We note that PGU has an advantage over the minimal $SU(5)$ grand unification (GUT): PGU provides a solution to the doublet-triplet splitting problem~\cite{Yanagida:1994vq,Witten:2001bf}.
The gauge coupling unification is still maintained approximately. 

We briefly discuss how non-universal gaugino masses arise in the $SU(5)_{\rm SM} \times SU(3)_H\times U(1)_H$ PGU model~\cite{Yanagida:1994vq},
where the unification of quarks and leptons into $SU(5)$ multiplets is maintained.
Gaugino masses are given by couplings between a SUSY breaking field $Z$ and gauge multiplets,
\begin{eqnarray}
\mathcal{L} &\supset& \int d^2 \theta \Bigr[ \Bigr( \frac{1}{4g_5^2} - \frac{k_5 Z}{M_P} \Bigl)W_5 W_5 + \Bigr( \frac{1}{4g_{3H}^2} - \frac{k_{3H} Z}{M_P} \Bigl)W_{3H} W_{3H} \nonumber \\
&&+  \Bigr( \frac{1}{4g_{1H}^2} - \frac{k_{1H} Z}{M_P} \Bigl)W_{1H} W_{1H} \Bigl] + {\rm h.c.} \,,
\end{eqnarray}
where $g_5$, $g_{3H}$ and $g_{1H}$ are the gauge coupling constants of $SU(5)_{\rm SM}$, $SU(3)_H$ and $U(1)_H$ gauge interactions, respectively. 
The field strength superfields of the gauge multiplets are denoted by $W_5$, $W_{3H}$ and $W_{1H}$, and 
$k_5$, $k_{3H}$ and $k_{1H}$ are constants.
After $SU(5)_{\rm SM} \times SU(3)_H\times U(1)_H$ is broken down to $SU(3)_C \times SU(2)_L \times U(1)_Y$, non-universal gaugino masses are generated at the GUT scale as
%
%
\begin{eqnarray}
M_1/M_2 \simeq \frac{k_5 \mathcal{N}+ k_{1H}}{k_5} \frac{1}{\mathcal{N}}, \ \ 
M_3/M_2 \simeq \frac{k_5 + k_{3H}}{k_5},
\end{eqnarray}
where we take the strong coupling limit, $g_{1H}^2, g_{3H}^2 \gg g_5^2$. 
The constant $\mathcal{N}$ is determined by the $U(1)_H$ charge of GUT breaking Higgs fields,
which break $SU(5)_{\rm SM} \times SU(3)_H\times U(1)_H$ down into the SM gauge group.
In the strong coupling limit of $SU(3)_H$ and $U(1)_H$, the gauge coupling unification is still maintained approximately as  $g_1^2 \simeq  g_3^2 \simeq g_2^2 = g_5^2$ at the GUT scale. Here, $g_1$, $g_2$ and $g_3$ are gauge coupling constants of $U(1)_Y$, $SU(2)_L$ and $SU(3)_C$, respectively.

\vspace{10pt}

\paragraph{i)FPUS}
The original focus point is proposed in a framework of gravity mediation. 
Surprisingly, if all the scalar masses are universal, their contributions to $\tilde m_{H}^2$ almost cancel each other at the low-energy scale~\cite{fp_original};
\begin{eqnarray}
&& \tilde m_{H}^2(3\,{\rm TeV}) \simeq  -1.170 M_3^2 + 0.072 m_{0}^2 + \dots,
\end{eqnarray}
where $\dots$ denotes other contributions containing $M_1$ or $M_2$.
If the ratio $m_0/M_3$ is fixed to be 4\,-\,5, 
the low-energy value of $\tilde m_{H}^2$ becomes insensitive to the SUSY breaking mass scale~\cite{Brummer:2013dya}.%
\footnote{
Originally, it is assumed that $M_3 \ll m_0$~\cite{fp_original}.
However, for the original scenario, the observed Higgs mass now pushes up the fine-tuning measure to $\Delta \sim 200$\,-\,$500$.
} 
(Due to the correction $\Delta V$ in Eq.~(\ref{eq:ewsb}), $\tilde m_{H}^2$ is not necessary negative for the successful EWSB.)

In {\bf FPUS} sleptons as well as squarks are as heavy as a few TeV to explain the observed Higgs mass; therefore  
the SUSY contribution to the muon $g-2$, $\Delta a_\mu$, is suppressed.



\paragraph{ii)FPGM}
In gaugino mediation models, we have a focus point with non-universal gaugino masses.
Assuming that scalar masses vanish at the GUT scale,
$\tilde m_{H}^2$ is given by
\begin{eqnarray}
\tilde m_{H}^2(3\,{\rm TeV}) \simeq -1.170 M_3^2 + 0.217 M_2^2 -0.109 M_2 M_3,
\end{eqnarray}
where we have dropped negligible contributions depending on $M_1$. 
One can see that above $\tilde m_{H}^2$ nearly vanishes for $M_2/M_3 \simeq  2.6$ and $-2.1$~\cite{fp_gaugino}. 
Universal scalar masses are introduced without much affecting the fine-tuning of the EWSB scale, as long as $m_0$ is not very large~\cite{fp_grav_nu_gaugino}.

Since $M_2$ is large, left-handed sleptons become inevitably heavy.
The low-energy value of $m_{L}^2$ is given by
\begin{eqnarray}
m_L^2(3\,{\rm TeV}) &\simeq& 0.391 M_2^2 + 0.033 M_1^2 +  ({\rm smaller\  terms}) \nonumber \\
&\simeq& 2.643 M_3^2 + 0.033 M_1^2  + ({\rm smaller\  terms}) \label{eq:mlsq},
\end{eqnarray}
where we take $M_2 = \, 2.6 M_3$ in the second line.
Consequently, {\bf FPGM} cannot explain $\Delta a_\mu \gtrsim 10^{-9}$.

\paragraph{iii)FPHSG}
It has been shown in Refs.~\cite{fp_gauge} that a focus point exists in high-scale gauge mediation models.%
\footnote{%
Although the gravitino mass $m_{3/2}$ is as large as $m_{3/2} \sim F_{\rm mess}/M_P$, it is assumed that
the contribution from gravity mediation is suppressed.
%
}
In {\bf FPHSG}, the number of $SU(2)_L$ doublet messengers ($N_L$) and $SU(3)_C$ triplet messengers ($N_D$) are not equal: for $N_L \gg N_D$, the EWSB scale becomes insensitive to the fundamental SUSY breaking parameter, $m_{\rm mess}$ (see Appendix \ref{sec:hsgmsb} for detail). 
\begin{eqnarray}
\tilde m_{H}^2(3\,{\rm TeV}) &\simeq& \frac{1}{N_D^2}\Bigl[ 0.217 N_L^2 -0.116 N_D N_L \nonumber \\
&+& 0.589 N_L - 1.175 N_D^2 -1.640 N_D \Bigr] M_3^2,
\end{eqnarray}
where $M_3 \simeq (\alpha_{\rm GUT}/(4\pi)) N_D m_{\rm mess}$.
For instance, $(N_L, N_D)=(29, 11)$ gives
\begin{eqnarray}
\tilde m_{H}^2(3\,{\rm TeV}) \simeq  0.017 M_3^2.
\end{eqnarray}
However, the masses of the wino and the mass squared of the left-handed slepton are proportional to $N_L$, and it is impossible to explain the discrepancy of the muon $g-2$.

\paragraph{iv)FPHGM}
We have a focus point in Higgs-gaugino mediation motivated by the $E_7$ non-linear sigma model~\cite{Kugo:1983ai}.
In Higgs-gaugino mediation, soft masses for squarks and sleptons vanish at $M_{\rm in}$, 
while those for the Higgs doublets are as large as gaugino masses.
This is consistent with non-observation of flavor-violating processes.
The low-energy $\tilde m_{H}^2$ is 
\begin{eqnarray}
\tilde m_{H}^2(3\,{\rm TeV}) \simeq  -1.167 M_{3}^2 + 0.693 m_H^2 + \dots \, .
\end{eqnarray}
Here, we assume that $m_{H_d}^2=m_{H_u}^2\equiv m_H^2$ at the high energy scale, for simplicity.
The ratio $m_{H}/M_3 \simeq 5/4$\,-\,$4/3$ leads to a small $\tilde m_{H}^2$~\cite{fp_hyy}.
In this model, sleptons as well as the wino can be light.
As is shown in the next section, it is possible to obtain $\Delta a_\mu \gtrsim 10^{-9}$.

\vspace{10pt}
As we have shown, among four focus point scenarios, only {\bf FPHGM} can explain the muon $g-2$ anomaly.
In the next section, we give a more detailed explanation for this point.

\section{The muon $g-2$ in the focus point SUSY}

The SUSY contribution to the muon $g-2$ is enhanced when gaugino(s) and smuon(s) are light. 
There are two dominant SUSY contributions to the muon $g-2$: 
wino-Higgsino-(muon sneutrino) diagram and bino-(L-smuon)-(R-smuon) diagram.
(Here, L and R denote left-handed and right-handed, respectively.) 
To enhance these contributions, at least, the left-handed slepton needs to be light.
Clearly, {\bf FPUS} can not explain the discrepancy of the muon $g-2$, since all the sleptons as well as squarks are heavy as a few TeV.
Also, L-smuon is too heavy to obtain $\Delta a_\mu \gtrsim 10^{-9}$ in {\bf FPGM} and {\bf FPHSG}. Therefore, only remaining possibility is {\bf FPHGM}.

The wino-Higgsino-(muon sneutrino) contribution to $(\Delta a_\mu)_{\rm SUSY}$ is given by~\cite{gm2_susy2}
\begin{eqnarray}
(a_\mu)_{\tilde{W}-{\tilde H}-{\tilde \nu}} &\simeq& (1- \delta_{2L}) \frac{\alpha_2}{4\pi}\frac{m_\mu^2 \tilde M_2 \mu}{m_{\tilde \nu}^4}\tan\beta \cdot 
F_C \left(\frac{\mu^2}{m_{\tilde \nu}^2}, \frac{\tilde M_2^2}{m_{{\tilde \nu}}^2} \right) \nonumber \\
&\simeq& 18.2 \times 10^{-10} \left(\frac{500 \, {\rm GeV}}{m_{\tilde \nu}}\right)^2 \frac{\tan\beta}{25}, 
\end{eqnarray}
where we take $\mu=(1/2) m_{\tilde \nu}$ and $\tilde M_2 = m_{\tilde \nu}$ in the second line.
Here, $\tilde M_2$ is the wino mass at the soft mass scale. 
The leading two-loop contribution $\delta_{2L}$ comes from large QED-logarithms~\cite{photonic, photonic_recent}
\begin{eqnarray}
\delta_{2L} =  \frac{4 \alpha}{\pi} \ln \frac{m_{\tilde \nu} }{m_\mu}.
\end{eqnarray}
To explain $\Delta a_\mu$ by this contribution, the masses of the wino and L-smuon should be around 500 GeV.
Obviously, the wino or L-smuon are too heavy to obtain $(a_\mu)_{\tilde{W}-{\tilde H}-{\tilde \nu}} \gtrsim 10^{-9}$ in {\bf FPUS}, {\bf FPGM} and {\bf FPHSG}.
In {\bf FPHGM}, on the other hand, the wino mass is unimportant for the focus point,
and hence can be small enough to explain the anomaly of the muon $g-2$.
As we will see, the L-smuon is also light enough.

The bino-(L-smuon)-(R-smuon) contribution is found to be~\cite{gm2_susy2}
\begin{eqnarray}
(a_\mu)_{\tilde{B}-{\tilde \mu}_L-{\tilde \mu}_R} &\simeq& 
(1- \delta_{2L}) \frac{3}{5}\frac{\alpha_1}{4\pi}\frac{m_\mu^2 \mu}{\tilde M_1^3}\tan\beta \cdot 
F_N \left(\frac{m_{{\tilde \mu}_L}^2}{\tilde M_1^2}, \frac{m_{{\tilde \mu}_R}^2}{\tilde M_1^2} \right) \nonumber \\
&\simeq& 21.7 \times 10^{-10} \frac{\mu}{640\, {\rm GeV}} \frac{\tan\beta}{40} \left(\frac{110\, {\rm GeV}}{\tilde M_1}\right)^3, \label{eq:gm2_bino}
\end{eqnarray}
where we take $m_{\tilde{\mu}_L}=3 \tilde M_1$ and $m_{\tilde{\mu}_R}=2 \tilde M_1$ in the second line. From the requirement of the small fine-tuning ($\Delta < 100$), there is an upper-bound on $\mu$: $\mu \lesssim 650$ GeV.
It can be seen that $(a_\mu)_{\tilde{B}-{\tilde \mu}_L-{\tilde \mu}_R}$ is sufficiently large only when the bino and smuons are very light as $200$-$300$ GeV,  and $\tan\beta$ is larger than 40. 
Although Eq.~(\ref{eq:gm2_bino}) does not contain $\tilde M_2$, it implicitly depends on $\tilde M_2$ through the renormalization group running from $M_{\rm in}$ to $M_{\rm IR}$: large $M_2$ thus $\tilde M_2$ leads to large L-slepton masses through the radiative corrections. Therefore, L-smuon becomes too heavy in {\bf FPGM} and {\bf FPHSG} (see Eq.(\ref{eq:mlsq})). 
Moreover, with large $\tan\beta \sim 40$, the tau Yukawa coupling becomes large and the stau mass becomes easily  tachyonic. Because of these reasons, it is difficult to obtain $(a_\mu)_{\tilde{B}-{\tilde \mu}_L-{\tilde \mu}_R} \gtrsim 10^{-9}$ in the known focus point SUSY scenarios.

\vspace{10pt}

In the following,
we discuss {\bf FPHGM} in detail.
We assume $M_1=M_3$, for simplicity.

\subsection{Focus point in Higgs-gaugino mediation (FPHGM)}

We consider the focus point in Higgs-gaugino mediation and estimate the fine-tuning of the EWSB scale in this model. For this purpose, we employ the following fine-tuning measure~\cite{ft_measure}:
\begin{eqnarray}
\label{eq:delta fpnus1}
\Delta &=& \max_a\{ |\Delta_a| \}, \nonumber \\
\Delta_a &=& \Bigl\{ \frac{\partial  \ln v}{\partial \ln \mu} \Bigr|_{v_{\rm obs}}, 
\frac{\partial  \ln v}{\partial \ln M_{3}} \Bigr|_{v_{\rm obs}}, 
\frac{\partial  \ln v}{\partial \ln M_{2}} \Bigr|_{v_{\rm obs}}, 
\frac{\partial  \ln v}{\partial \ln B_0} \Bigr|_{v_{\rm obs}} \Bigr\},  \label{eq:ft_hgm}
\end{eqnarray}
where $v_{\rm obs} \simeq 174.1$ GeV. The fundamental mass parameters in $\Delta_a$ are defined at $M_{\rm in}=10^{16}$ GeV. 
As shown in Eq.(\ref{eq:ewsb}), 
The VEV $v$ in $\Delta_a$ is determined by the Higgs potential including one-loop radiative corrections, which are in fact non-negligible.
It is very interesting if there is a small $\Delta$ region where the observed Higgs boson mass and the muon anomaly $g-2$ are simultaneously explained.

In our numerical calculations, the Higgs boson mass is calculated using {\tt FeynHiggs 2.10.3}~\cite{feynhiggs} and the SUSY mass spectra as well as $\Delta$ is evaluated utilizing {\tt SoftSUSY 3.6.1}~\cite{softsusy}. The strong coupling constant and the top pole mass are taken as $\alpha_s(M_Z)=0.1185$ and $m_{t}=173.34$ GeV.

\begin{figure}[t]
\begin{center}
\includegraphics[scale=1.25]{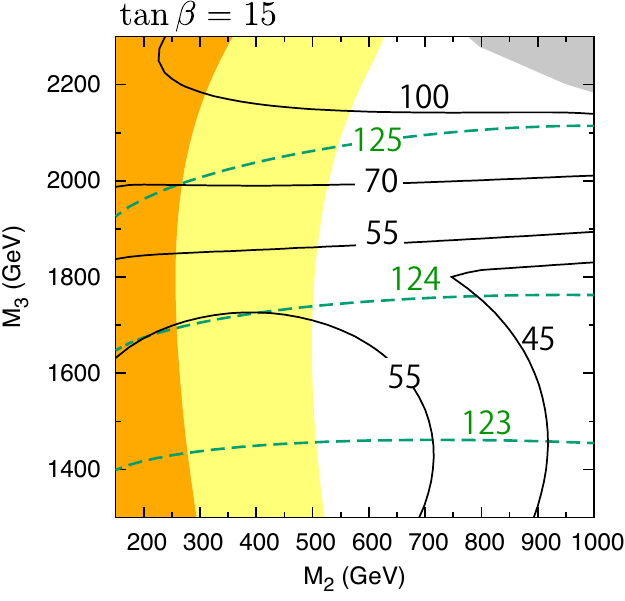}
\includegraphics[scale=1.25]{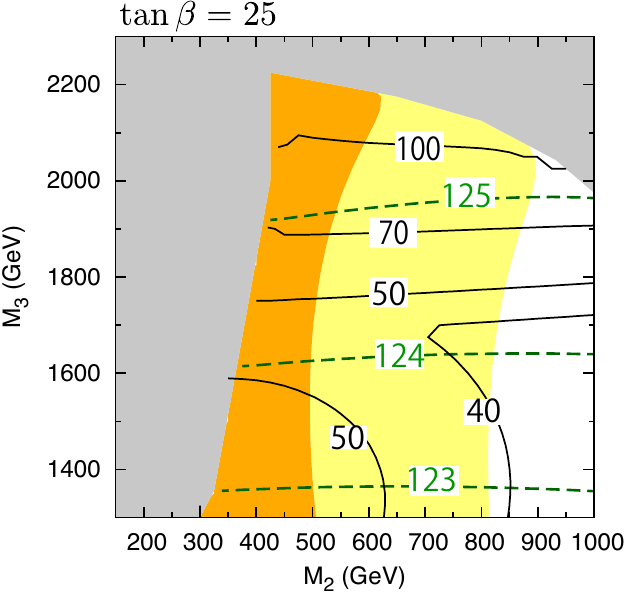}
\caption{The contours of $\Delta$ (black solid line) and $m_h$ (green dashed line) in {\bf FPHGM}. 
The Higgs mass $m_{h}$ is shown in the unit of GeV. We take $ m_H/M_3 = 4/3$. 
In the orange (yellow) region the SUSY contribution to the muon $g-2$ reduces the discrepancy to 1$\sigma$ (2$\sigma$).
The gray regions are excluded since the stau becomes too light (left part, $m_{ {\tilde \tau}_1 }$ or $m_{ {\tilde \nu}_{\tau} }<100$ GeV) or the EWSB does not occur (upper part). 
}
\label{fig:hyy_cont}
\end{center}
\end{figure}

We show the contours for the Higgs boson mass and $\Delta$ in Fig.~\ref{fig:hyy_cont}. In the orange (yellow) region, the SUSY contribution $\Delta a_\mu$ reduces the discrepancy of the muon $g-2$ from the SM prediction to 1$\sigma$ (2$\sigma$).
For the SM prediction of the muon $g-2$, we use $(a_\mu)_{\rm EXP}-(a_\mu)_{\rm SM} = (26.1 \pm 8.0) \cdot 10^{-10}$ (see Eq.(\ref{eq:gm2_dis})).
The gray regions are excluded since the stau becomes too light (left part) or the EWSB does not occur (upper part).
For $\tan\beta=15$ (25), $M_2$ smaller than 300 (500) GeV can reduce the discrepancy of the muon $g-2$ to 1$\sigma$ level. 
Here, $M_2=(300, 500)$\,GeV corresponds to the wino mass around (200, 370)\,GeV at the stop mass scale $M_{\rm IR}$. The observed Higgs boson mass around 125 GeV is also consistently explained with $\Delta=40$\,-\,$100$.

\begin{figure}[t]
\begin{center}
\includegraphics[scale=1.70]{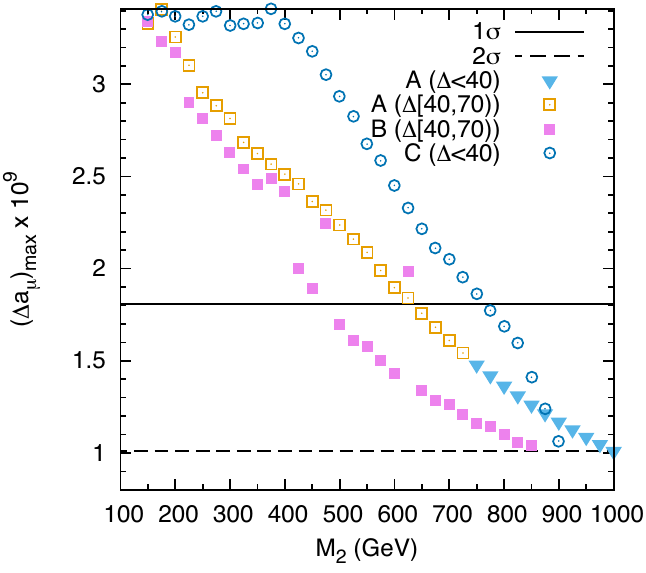}
\caption{The maximum value of $\Delta a_\mu \times 10^{9}$ in {\bf FPHGM} for different parameter sets. Here, A:$\ (M_3,  m_H/M_3 )=(1500, 4/3)$, B:$\ (1900, 4/3)$, C:$\ (1400, 1.37)$. In each point, $\tan\beta$ is varied within a range $[10:60]$, requiring $m_{ {\tilde \tau}_1 }, m_{ {\tilde \nu}_{\tau} } > 100$ GeV.
}
\label{fig:amu_hyy}
\end{center}
\end{figure}

Also, we show the maximum value of $\Delta a_\mu$ in Fig.~\ref{fig:amu_hyy} for different parameter sets (A, B, C).
We vary $\tan\beta$ within a range $[10:60]$ in each parameter set such that $\Delta a_\mu$ is maximized.
We require that $m_{{\tilde \tau}_1}$, $m_{ {\tilde \nu}_{\tau} } \gtrsim 100$ GeV; therefore  
 the region with too small $m_{L}$ or too large $\tan\beta$ is not allowed. (The allowed range of $\tan\beta$ is up to $\sim$ 30 in the parameter region preferred for the muon $g-2$.)
The maximum value of $\Delta a_\mu$ easily exceeds $1.8 \cdot 10^{-9}$ in the mild fine-tuning region. 
For C, $M_2$ smaller than 750 GeV (580 GeV at $M_{\rm IR}$) is allowed to explain the anomaly of the muon $g-2$. In this case, the level of the fine-tuning is still as low as $\Delta<40$.

Interestingly, in this {\bf FPHGM} the muon $g-2$ anomaly is easily explained. This is due to the smallness of scalar masses at $M_{\rm in}$, which gives small radiative corrections to the staus during the RGE running: the lighter L-smuon and larger $\tan\beta$ are allowed compared to models which will be discussed in the next section.

Let us present some sample mass spectra and $\Delta$ in Table \ref{table:fphgm}. 
One can see that the discrepancy of the muon $g-2$ is, in fact, explained in the region $\Delta \sim 40$\,-\,$80$. The calculated Higgs boson mass is consistent with the observed value. 
Note that the tau sneutrino is the lightest SUSY particle (LSP) in these model points, and one may need to pay attention to it.

\subsection{Sneutrino LSP}
\label{sec:sneutrino LSP}
Before closing this section, let us comment on the (tau) sneutrino LSP from view points of the cosmology and collider searches, since the tau sneutrino tends to be the LSP in the parameter region of our interest (apart from the region where the wino mass is around 100 GeV).
If the sneutrino LSP is absolutely stable, it is easily excluded by direct detection experiments due to a large scattering cross section with nuclei~\cite{Falk:1994es}. 
However, the sneutrino LSP can easily decay into SM particles with a life-time less than $0.1$\,-\,$1$ sec, if there is a tiny R-parity violation (e.g. $W=LL \bar E, LQ \bar D$). Therefore, the sneutrino LSP neither conflicts with the direct detection experiments nor standard cosmology.

The sneutrino LSP may behave as a stable particle inside the detector.
In this case,
the sneutrino can be searched for at the LHC through the production of chargino-neutralino, which eventually decay into multi-leptons with a missing transverse momentum.
It may be distinguishable from an ordinary neutralino LSP case, since the flavors of the final state leptons are uncorrelated for the sneutrino LSP~\cite{sneutrino_lsp_lhc}.

\begin{table}[t!]
   \caption{
   Model points of {\bf FPHGM}. Here, $M_1=M_3$ at $M_{\rm in} (= 10^{16}$ GeV) is assumed .
   }
  \begin{center}
  \small
    \begin{tabular}{  c | c  }
            {\bf P1} & \\
\hline
    $M_{3} $ &  2000 GeV \\
    $M_{2} $ &  400 GeV \\
    $ m_H/M_3 $ &  4/3\\
    $\tan \beta$ &  20\\
    \hline
\hline
    $\mu$ & 353 \\
    $\Delta$ & 82 \\    
    $m_{\rm gluino}$ & 4.14  TeV \\
      $m_{\tilde{q}}$ & 3.55\,-\,3.56 TeV \\
    $m_{\tilde{t}_{1,2}}$ & 2.82, 3.16 TeV \\
    $m_{\tilde{\mu}_L}$, $m_{\tilde{\mu}_{R}}$ & 403 GeV, 739 GeV\\
    $m_{\tilde{\tau}_1}$, $m_{\tilde{\nu}_{\tau}}$ & 271 GeV, 261 GeV\\
     $m_{\chi_1^0}$, $m_{\chi_2^0}$ & 267, 366 GeV \\
     $m_{\chi_3^0}$, $m_{\chi_4^0}$ & 395, 876 GeV \\
     $m_{\chi_1^{\pm}}$, $m_{\chi_2^{\pm}}$ & 269, 401 GeV \\
     $m_{h}$ & 125.2 GeV \\
     $\Delta a_\mu$ & 19.2 $\cdot \, 10^{-10}$ \\
    \end{tabular}
        \hspace{2pt}
    \begin{tabular}{  c | c  }
            {\bf P2} & \\
\hline
    $M_{3} $ & 1650 GeV \\
    $M_{2} $ & 495 GeV \\
    $ m_H/M_3 $ & 4/3 \\
    $\tan \beta$ & 27 \\
    \hline
\hline
    $\mu$ & 433 \\
    $\Delta$ & 45 \\    
    $m_{\rm gluino}$ & 3.46  TeV \\
      $m_{\tilde{q}}$ & 2.98\,-\,3.00 TeV \\
    $m_{\tilde{t}_{1,2}}$ & 2.35, 2.64 TeV \\
    $m_{\tilde{\mu}_L}$, $m_{\tilde{\mu}_{R}}$ & 408 GeV, 611 GeV\\
    $m_{\tilde{\tau}_1}$, $m_{\tilde{\nu}_{\tau}}$ & 211 GeV, 208 GeV\\
     $m_{\chi_1^0}$, $m_{\chi_2^0}$ & 350, 444 GeV \\
     $m_{\chi_3^0}$, $m_{\chi_4^0}$ & 474, 721 GeV \\
     $m_{\chi_1^{\pm}}$, $m_{\chi_2^{\pm}}$ & 352, 480 GeV \\
     $m_{h}$ & 124.1 GeV \\
     $\Delta a_\mu$ & 19.9 $\cdot \, 10^{-10}$ \\
    \end{tabular}    
  \label{table:fphgm}
  \end{center}
\end{table}

\section{Variants of FPUS} \label{sec:fpnus}
So far, among known focus point SUSY scenarios, only {\bf FPHGM} can explain the muon $g-2$ anomaly. In this section, we discuss possible modifications of other focus point SUSY scenarios. 

In  {\bf FPGM} and {\bf FPHSG}, the heavy wino is crucial for realizing semi-natural SUSY; therefore, it is very difficult to modify these scenarios to be consistent with the muon $g-2$ experiment. 
On the other hand, the modification may be possible for {\bf FPUS} by relaxing the condition of universal scalar masses and taking  slepton masses to be small.
Although the fine-tuning is rather insensitive to the slepton masses, 
this modification is not very easy.
This is because
radiative corrections induce negative squared masses for staus. 
Staus become very light or tachyonic via radiative corrections for $\tan\beta=\mathcal{O}(10)$.
We have found, however, two possible modifications of {\bf FPUS}, which we refer to as {\bf FPNUS1} and {\bf FPNUS2}. 
Here, {\bf FPNUS1} respects the $SU(5)$ unification while {\bf FPNUS2} does not.%
\footnote{
Here, the ``$SU(5)$ unification" means the unification quarks and leptons into $SU(5)$ multiplets.
The GUT gauge group itself is not assumed to be a single $SU(5)$. See the comment on product groups in section~\ref{sec:focus EWSB}.
}
In {\bf FPNUS1}, the discrepancy of the muon $g-2$ from the SM prediction is reduced to 1$\sigma$ level
for the wino as light as $\sim 100$ GeV when $m_{H_u}\sim m_{H_d}$ at $M_{\rm in}$,
and for the wino as light as $\sim 400$ GeV when $m_{H_d} \ll m_{H_u}$ at $M_{\rm in}$.
There is a larger parameter space in {\bf FPNUS2}.

\paragraph{v)FPNUS1}
Scalar contributions to $m_{H_u}^2$ can be cancelled, even if scalar masses are not completely universal.
Similar to {\bf FPUS}, we take $m_{Q}=m_{\bar U}=m_H=m_{\bar E}$.
Then, we have 
\begin{eqnarray}
\tilde m_{H}^2(3\,{\rm TeV}) &\simeq&  -1.170 M_3^2 + 0.069 m_Q^2 + 0.074 m_{L}^2 -0.070 m_{\bar D}^2  + \dots \, .
\end{eqnarray}
Note that the relation $m_{Q}=m_{\bar U}=m_{\bar E}$ is consistent with the $SU(5)$ unification.
In the $SU(5)$ unification, the relation $m_{\bar D}=m_{L}$ is imposed, which we take as a free parameter independent of $m_Q$.
Assuming that $m_Q/M_3 \sim 4$\,-\,$5$, we obtain the focus point.

\paragraph{vi)FPNUS2}
There is another focus point once the $SU(5)$ unification is abandoned.
For $m_{Q}=m_{\bar U}=m_{\bar D}=m_{H}$ with a fixed ratio of $m_Q/M_3$, small $\tilde m_H^2$ compared to  $M_3^2$ can be obtained as well, although this condition is not consistent with the $SU(5)$ unification.

\subsection{FPNUS1}
Let us evaluate the fine-tuning $\Delta$, the Higgs boson mass and $\Delta a_\mu$ in {\bf FPNUS1}.
Here, we consider the case of $m_{Q}=m_{U}=m_{E}=m_{H}$ and the fixed ratio $m_{Q}/M_{3}$. 
Also, $m_{L}=m_{\bar D}$ is assumed so that quarks and leptons are unified into $SU(5)$ multiplets.
The fine-tuning of this model can be estimated by the following measure: 
\begin{eqnarray}
\label{eq:delta fpnus1}
\Delta &=& \max_a\{ |\Delta_a| \}, \nonumber \\
\Delta_a &=& \Bigl\{ \frac{\partial  \ln v}{\partial \ln \mu} \Bigr|_{v_{\rm obs}}, 
\frac{\partial  \ln v}{\partial \ln M_{3}} \Bigr|_{v_{\rm obs}}, 
\frac{\partial  \ln v}{\partial \ln M_{2}} \Bigr|_{v_{\rm obs}}, 
\frac{\partial  \ln v}{\partial \ln m_L} \Bigr|_{v_{\rm obs}}, 
\frac{\partial  \ln v}{\partial \ln B_0} \Bigr|_{v_{\rm obs}} \Bigr\}.
\end{eqnarray}

\begin{figure}[t]
\begin{center}
\includegraphics[scale=1.00]{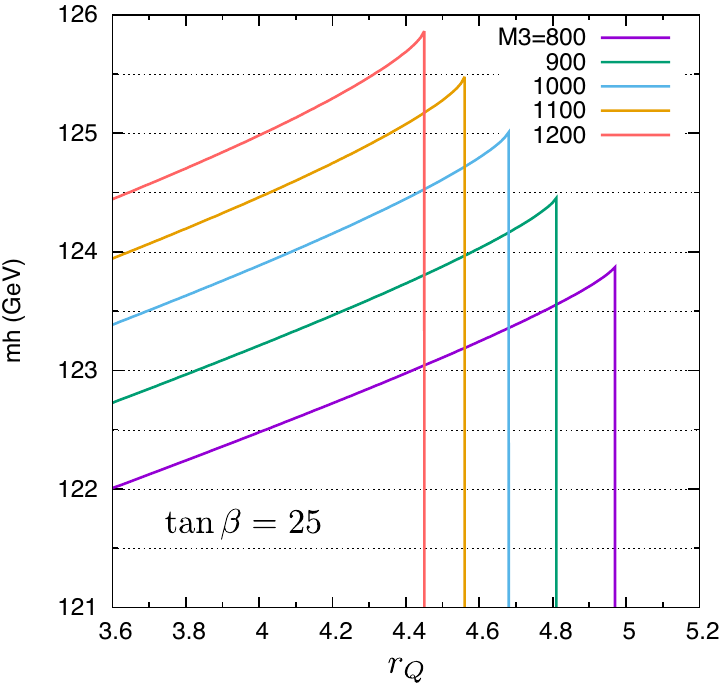}
\includegraphics[scale=1.00]{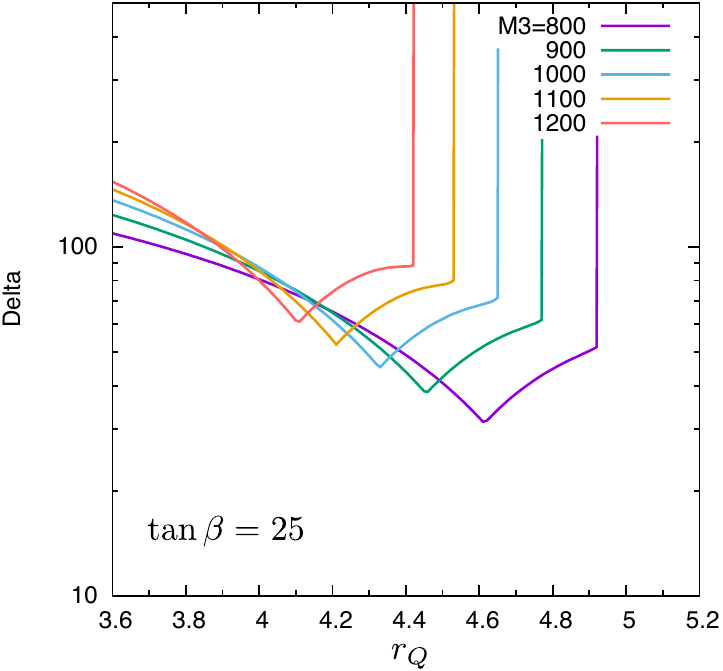}
\includegraphics[scale=1.00]{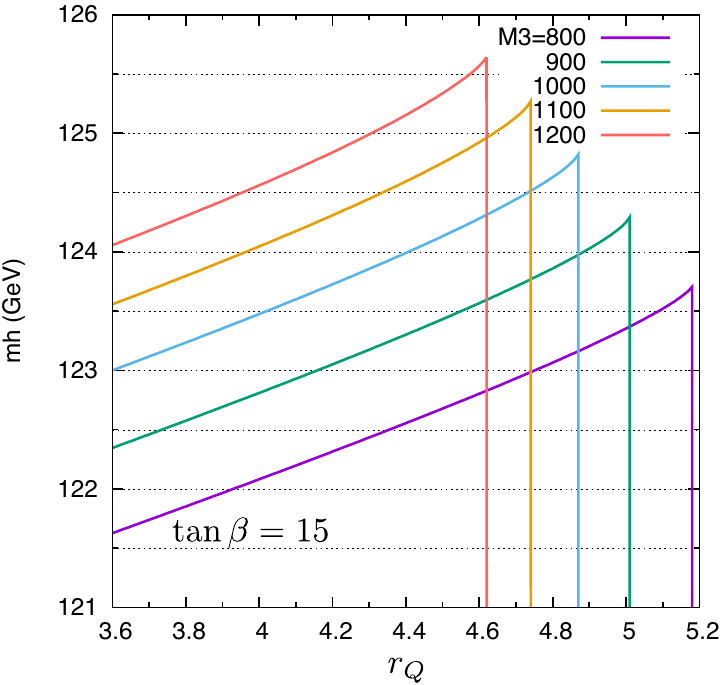}
\includegraphics[scale=1.00]{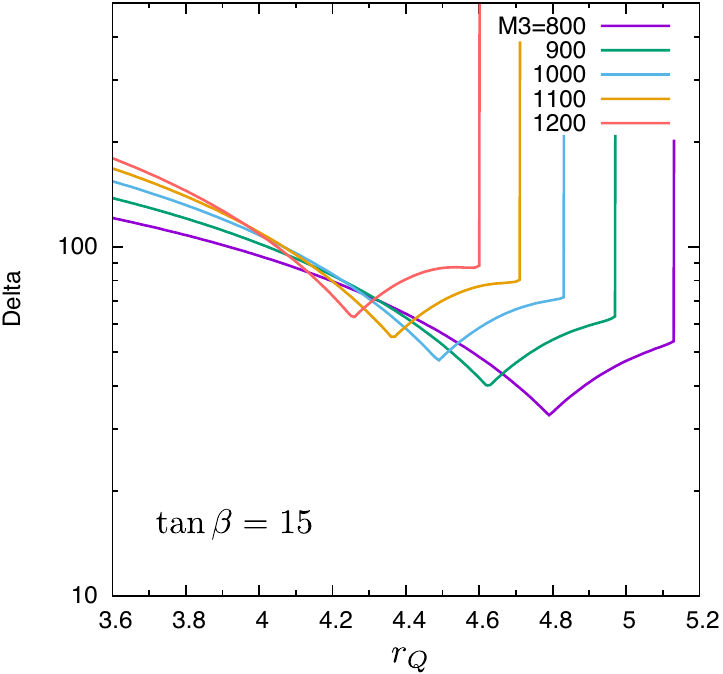}
\caption{The Higgs boson mass and $\Delta$ in {\bf FPNUS1}, 
with parameter sets $(M_3, M_2, m_L)=(800$-$900,500, 1000), (1000$-$1200,500, 1200) $ GeV.
In the upper (lower) panel, $\tan\beta=25$ (15).
Here, $r_Q \equiv m_Q/ M_3$.
}
\label{fig:fig_mh_delta}
\end{center}
\end{figure}

In Fig.~\ref{fig:fig_mh_delta}, the Higgs boson mass $m_h$ and $\Delta$ are shown for different $M_3$. 
Here, $r_Q$ is the ratio of the squark mass to the gluino mass, $m_{Q}/M_3$. The gluino mass at $M_{\rm in}$ is taken as $M_{3}=(800, 900, 1000, 1100, 1200)$ GeV. 
As $r_Q$ increases,
$\Delta$ is minimized at a certain point.
Above the vertical line,
the EWSB no longer occurs.
%
%
In small $\Delta$ region, the calculated Higgs boson mass of $m_h \simeq (123.5, 124.5, 125)$ GeV is   obtained for $M_3=(800, 900, 1000)$ GeV and $\tan\beta=25$, while larger $M_3$ is required for $\tan\beta=15$.

\begin{figure}[t]
\begin{center}
\includegraphics[scale=1.70]{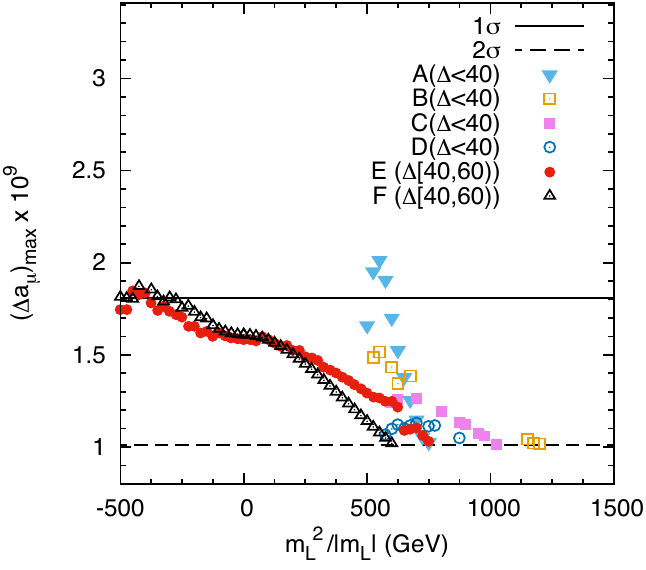}
\caption{The maximum value of $\Delta a_\mu \times 10^{9}$ in {\bf FPNUS1} for different parameter sets.
A:$\ (M_3, M_2, m_Q/M_3, m_{H_d}/m_Q)=(750, 150, 5.5, 1.0)$, 
B:$\ (750, 400, 5.1, 1.0)$, 
C:$\ (900, 150, 4.7, 1.0)$, 
D:$\ (900, 400, 4.6, 1.0)$,
E:$\ (900, 150, 5.6, 0.3)$,
F:$\ (750, 400, 6.0, 0.2)$.
In each point, $\tan\beta$ is varied within a range $[10:60]$, requiring $m_{ {\tilde \tau}_1 }, m_{ {\tilde \nu}_{\tau} } > 100$ GeV.
The condition $m_{H_d}=m_{H_u}$ is relaxed for E and F.
}
\label{fig:fig_fpnus1}
\end{center}
\end{figure}

Next, we see whether we can explain the muon $g-2$ anomaly in {\bf FPNUS1}.  
In Fig.~\ref{fig:fig_fpnus1}, the maximum value of $\Delta a_\mu$ in the region with mild fine-tuning is shown.
We take different parameter sets denoted by A, B, C, D, E and F as shown in the caption. We vary $\tan\beta$ within a range $[10:60]$ in each parameter set such that $\Delta a_\mu$ is maximized. (The allowed range of $\tan\beta$ is up to $\sim$ 20.)
%
%
We see that, in the very light wino case A, the discrepancy of the muon $g-2$ from the SM prediction can be reduced to 1$\sigma$ level, while in the heavier wino case B the discrepancy is reduced to 1.5$\sigma$. In E and F, the condition $m_{H_u}=m_{H_d}$ at $M_{\rm in}$ is relaxed, and there is a region where the discrepancy is reduced to $1\sigma$ level for $m_L^2(M_{\rm in}) <0$.
Note that $m_{H_u}\neq m_{H_d}$ is consistent with the $SU(5)$ unification.

Let us present a sample mass spectrum and $\Delta$ in Table \ref{table:fpnus} ({\bf P3}). 
One can see the discrepancy of the muon $g-2$ is reduced around 1$\sigma$ if the wino-like chargino is as light as $\sim 100$ GeV.

\subsection{FPNUS2}
Once we abandon the $SU(5)$ unification, we have another focus point ({\bf FPNU2}).  
Here, we consider the case for $m_{Q}=m_{\bar U}=m_{\bar D}=m_{H}$ with the fixed ratio of $m_Q/M_3\equiv r_Q$. 
Although this model is not consistent with the $SU(5)$ unification, a larger parameter space with $\Delta a_\mu \gtrsim 1.8 \cdot  10^{-9}$  exists.
The fine-tuning measure $\Delta$ is slightly changed from {\bf FPNU1} as~\footnote{
Unless $m_{L}$ or $m_{\bar E}$ is very large, $\Delta$ is dominated by $\Delta_\mu$ or $\Delta_{M_3}$ so far.
}
\begin{eqnarray}
\label{eq:delta fpnus2}
\Delta &=& \max_a\{ |\Delta_a| \},  \\
\Delta_a &=& \Bigl\{ \frac{\partial  \ln v}{\partial \ln \mu} \Bigr|_{v_{\rm obs}},
 \frac{\partial  \ln v}{\partial \ln M_{3}} \Bigr|_{v_{\rm obs}}, 
  \frac{\partial  \ln v}{\partial \ln M_{2}} \Bigr|_{v_{\rm obs}}, 
  \frac{\partial  \ln v}{\partial \ln m_L} \Bigr|_{v_{\rm obs}}, 
    \frac{\partial  \ln v}{\partial \ln m_{\bar E}} \Bigr|_{v_{\rm obs}}, 
 \frac{\partial  \ln v}{\partial \ln B_0} \Bigr|_{v_{\rm obs}} \Bigr\}.  \nonumber 
\end{eqnarray}

\begin{figure}[t]
\begin{center}
\includegraphics[scale=1.70]{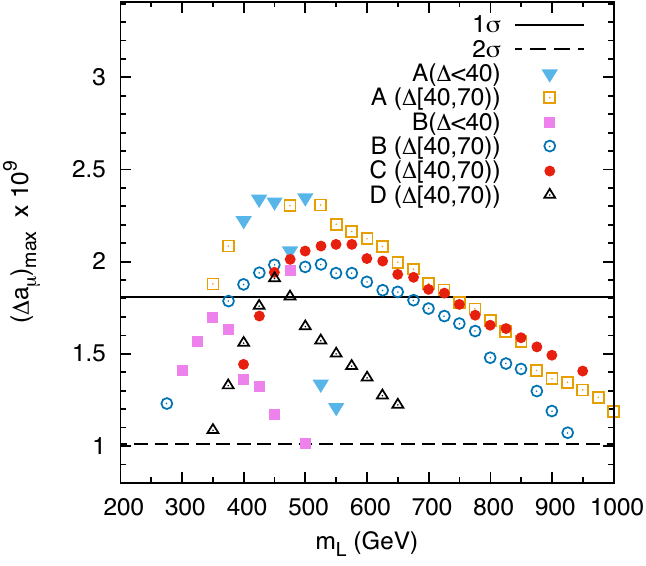}
\caption{The maximum value of $\Delta a_\mu \times 10^{9}$ in {\bf FPNUS2} for different parameter sets. 
A:$\ (M_3, M_2, m_Q/M_3, m_{\bar{E}}/m_L)=(750, 150, 5.1, 2.0)$, B:$\ (750, 400, 5.1, 2.0)$, C:$\ (900, 150, 5.0, 2.0)$, D:$\ (900, 400, 5.1, 2.0)$. In each points, $\tan\beta$ is  is varied within a range $[10:60]$.
}
\label{fig:fig_fpnus2}
\end{center}
\end{figure}

In Fig.~\ref{fig:fig_fpnus2}, the maximum value of $\Delta a_\mu$ for different parameter sets are shown. Here, we only consider the mild fine-tuning region. (The Higgs boson mass is similar to the one in {\bf FPNUS1}.)
As in the case of {\bf FPNUS1}, $\tan\beta$ is varied within a range $[10:60]$ to find a maximum value of $\Delta a_\mu$. One can see that A and B can reduce the discrepancy of the muon $g-2$ to 1$\sigma$ level with $\Delta <40$. If the required upper-bound on $\Delta$ is relaxed to $\Delta < 70$, all parameter sets (A, B, C, D) shown in the figure can reduce the discrepancy of the muon $g-2$ to 1$\sigma$ level: the Higgs boson mass and the muon $g-2$ anomaly are explained relatively easily in  {\bf FPNUS2} compared to {\bf FPNUS1}. This is because the stau is heavier for the same L-slepton mass at $M_{\rm IR}$ and $\tan\beta$.
This allows larger $\tan\beta$ and smaller $m_{\tilde L}$, avoiding the too light stau.
%

%

Finally, let us present a sample mass spectrum and $\Delta$ in Table \ref{table:fpnus} ({\bf P4}). Although this model is not consistent with the $SU(5)$ unification, the anomaly of the muon $g-2$ is, in fact, explained in the region with $\Delta \simeq 60$.

\begin{table}[t!]
   \caption{
 Model points of {\bf FPNUS1} ({\bf P3}) and {\bf FPNUS2} ({\bf P4}) are shown.
Here, $M_1=M_3$ and $m_{H_d}=m_{H_u}$ at $M_{\rm in} (= 10^{16}$ GeV) is assumed.
   }
  \begin{center}
  \small
    \begin{tabular}{  c | c  }
            {\bf P3} & \\
\hline
    $M_{3} $ &  800 GeV \\
    $M_{2} $ &  200 GeV \\
    $-$ & \\
    $m_{L}=m_{\bar D} $ & 560 GeV \\
    $ m_Q/M_3$ & 5.3 \\
    $\tan \beta$ & 13 \\
    \hline
\hline
    $\mu$ & 221 \\
    $\Delta$ & 40 \\    
    $m_{\rm gluino}$ & 1.89 TeV \\
      $m_{\tilde{q}}$ & 1.46\,-\,4.46 TeV \\
    $m_{\tilde{t}_{1,2}}$ & 2.85, 3.71 TeV \\
     $m_{\tilde{\mu}_L}$, $m_{\tilde{\mu}_{R}}$ & 435 GeV, 4251 GeV\\
    $m_{\tilde{\tau}_1}$, $m_{\tilde{\nu}_{\tau}}$ & 160 GeV, 139 GeV\\
     $m_{\chi_1^0}$, $m_{\chi_2^0}$ & 126, 236 GeV \\
     $m_{\chi_3^0}$, $m_{\chi_4^0}$ & 254, 364 GeV \\
     $m_{\chi_1^{\pm}}$, $m_{\chi_2^{\pm}}$ & 129, 269 GeV \\
     $m_{h}$ & 123.8 GeV \\
     $\Delta a_\mu$ & 17.5 $\cdot \, 10^{-10}$ \\
     & \\
    \end{tabular}
    \begin{tabular}{  c | c  }
            {\bf P4} & \\
\hline
    $M_{3} $ &  1000 GeV \\
    $M_{2} $ &  350 GeV \\
    $m_{\bar E} $ & 1000 GeV \\
    $m_{\bar L} $ &  560 GeV \\
    $ m_Q/M_3 $ & 4.9 \\
    $\tan \beta$ & 19 \\
    \hline
\hline
    $\mu$ & 168 \\
    $\Delta$ & 62 \\    
    $m_{\rm gluino}$ & 2.37 TeV \\
      $m_{\tilde{q}}$ & 5.16\,-\,5.18 TeV \\
    $m_{\tilde{t}_{1,2}}$ & 3.34, 4.25 TeV \\
     $m_{\tilde{\mu}_L}$, $m_{\tilde{\mu}_{R}}$ & 515 GeV, 984 GeV\\
    $m_{\tilde{\tau}_1}$, $m_{\tilde{\nu}_{\tau}}$ & 143 GeV, 119 GeV\\
     $m_{\chi_1^0}$, $m_{\chi_2^0}$ & 146, 181 GeV \\
     $m_{\chi_3^0}$, $m_{\chi_4^0}$ & 310, 445 GeV \\
     $m_{\chi_1^{\pm}}$, $m_{\chi_2^{\pm}}$ & 154, 314 GeV \\
     $m_{h}$ & 125.0 GeV \\
      $\Delta a_\mu$ & 18.6 $\cdot \, 10^{-10}$\\
     & \\
    \end{tabular}

  \label{table:fpnus}
  \end{center}
\end{table}

\section{Conclusions}

The focus point SUSY scenario is very attractive, since it explains semi-naturally
the observed electroweak breaking scale $v\simeq 174.1$ GeV even when masses of squarks and gluino are
in several TeV region. One interesting prediction of the focus point SUSY breaking scenario
is the light Higgsino with a mass of several hundred GeV. This relatively light Higgsino
provides a possibility of explaining the anomaly of the muon $g-2$.
In fact, if the wino and the left-handed smuon are also light, the anomaly of the muon $g-2$ is explained.

In this paper,
we have found that, among the known focus point SUSY scenarios, a scenario based on the Higgs-gaugino mediation can explain the observed value of the $g-2$ with mild fine-tuning measures $\Delta =40$\,-\,$80$. This scenario is proposed recently by the current authors motivated by $E_7$ non-linear sigma model, which may
explain why the family number is three.
There, the wino mass is unimportant for the focus point and hence can be light enough.
The mass of the left-handed smuon is mainly given by the quantum correction from the wino loop and is small.

The tau-sneutrino is likely to be the LSP in the parameter region of our interest,  which gives a distinctive collider signal as described in Sec.~\ref{sec:sneutrino LSP}. Therefore, this intriguing possibility may be tested and distinguished from other SUSY scenarios at the LHC.

Also, we propose two new focus point SUSY scenarios based on gravity mediation, which are variants of the well known focus point SUSY scenario. Unlike the original one, the scalar masses are no longer universal and the left-handed sleptons are light.
We have shown that the muon $g-2$ anomaly is explained.


In this paper, we have mainly discussed the anomaly of the muon $g-2$ in
focus point SUSY scenarios.
The focus point SUSY needs some relations among relevant mass parameters.
We hope that those relations may be given by more fundamental physics (see e.g.~\cite{Brummer:2013dya,fp_model}).
It is, however, beyond the scope of this paper.

\section*{Acknowledgments}
This work is supported by Grant-in-Aid for Scientific research from
the Ministry of Education, Culture, Sports, Science
and Technology (MEXT) in Japan,
No.\ 26104009 and 26287039 (T.\,T.\,Y.),
and also by World Premier International Research Center Initiative (WPI Initiative), MEXT, Japan (T.\,T.\,Y.).
The research leading to these results has received funding
from the European Research Council under the European Unions Seventh
Framework Programme (FP/2007-2013) / ERC Grant Agreement n. 279972
``NPFlavour'' (N.\,Y.).
The work of K.\,H.\, is supported in part by a JSPS Research Fellowships for Young Scientists.

\appendix

\section{High scale gauge mediation} \label{sec:hsgmsb}
We consider a high-scale gauge mediation model with $N_L$ pairs of $SU(2)_L$ doublet messengers and $N_D$ pairs of $SU(3)_C$ triplet messengers. The SUSY breaking mass and SUSY invariant mass of the messenger superfield are denoted by $F_{\rm mess}$ and $M_{\rm mess}$, respectively.  It is assumed that $F_{\rm mess}$ and $M_{\rm mess}$ are common for all the messenger fields.

In this setup, the gaugino masses are given by
\begin{eqnarray}
M_1 = \frac{\alpha_1}{4\pi} m_{\rm mess} \left(\frac{3}{5} N_L + \frac{2}{5} N_D\right), \ M_2 = \frac{\alpha_2}{4\pi}m_{\rm mess} N_L, \ \ M_3 = \frac{\alpha_3}{4\pi} m_{\rm mess} N_D,
\end{eqnarray}
where $m_{\rm mess}=F_{\rm mess}/M_{\rm mess}$. The scalar masses are 
\begin{eqnarray}
m_{Q}^2 &=&  \Bigl[ \frac{8}{3}\left(\frac{\alpha_3}{4\pi} \right)^2 N_D 
+\frac{3}{2}\left(\frac{\alpha_2}{4\pi} \right)^2 N_L 
+ \frac{1}{30}\left(\frac{\alpha_1}{4\pi} \right)^2 \left( \frac{3}{5}N_L + \frac{2}{5}N_D\right)
 \Bigr] m_{\rm mess}^2 , \nonumber \\
m_{\bar U}^2 &=& \Bigl[ \frac{8}{3}\left(\frac{\alpha_3}{4\pi} \right)^2 N_D  
+ \frac{8}{15}\left(\frac{\alpha_1}{4\pi} \right)^2 \left( \frac{3}{5}N_L + \frac{2}{5}N_D\right)
\Bigr] m_{\rm mess}^2,  \nonumber \\
m_{\bar D}^2 &=&  \Bigl[ \frac{8}{3}\left(\frac{\alpha_3}{4\pi} \right)^2 N_D  
+\frac{2}{15}\left(\frac{\alpha_1}{4\pi} \right)^2 \left( \frac{3}{5}N_L + \frac{2}{5}N_D\right)
\Bigr] m_{\rm mess}^2 ,  \nonumber \\
m_{L}^2 &=&  \Bigl[ \frac{3}{2}\left(\frac{\alpha_2}{4\pi} \right)^2 N_L  
+\frac{3}{10}\left(\frac{\alpha_1}{4\pi} \right)^2 \left( \frac{3}{5}N_L + \frac{2}{5}N_D\right)
\Bigr] m_{\rm mess}^2 , \nonumber \\
m_{\bar E}^2 &=&  \Bigl[ \frac{6}{5}\left(\frac{\alpha_1}{4\pi} \right)^2 \left( \frac{3}{5}N_L + \frac{2}{5}N_D\right) \Bigr] m_{\rm mess}^2 , \nonumber \\
m_{H_u}^2 &=& m_{H_d^2}= m_{L}^2.
\end{eqnarray}

If we take $M_{\rm mess}=M_{\rm GUT}$, the low-energy value of $m_{H_u}^2-(m_{H_d}^2-m_{H_u}^2)/\tan^2\beta$ $(\equiv \tilde m_H^2)$  is  written by
\begin{eqnarray}
\tilde m_{H}^2(3\,{\rm TeV}) &\simeq& \frac{1}{N_D^2}\Bigl[ 0.216 N_L^2 -0.116 N_D N_L \nonumber \\
&+& 0.587 N_L - 1.172 N_D^2 -1.636 N_D \Bigr] M_3^2,
\end{eqnarray}
where we take $\tan\beta=20$.

\end{document}